\documentclass[aps,twocolumn,prl,preprintnumbers,amsmath,amssymb]{revtex4}
\usepackage{graphicx, bm}
\usepackage{dcolumn}
\usepackage{float}
\usepackage{latexsym}
\usepackage{amsmath}
\usepackage{graphics}
\usepackage{amssymb}
\usepackage{layout}
\usepackage{verbatim}
\usepackage{amsfonts,epsfig}
\usepackage{color}
\usepackage{setspace}
\newcommand{\beqnar}{\begin{eqnarray}}
\newcommand{\eeqnar}{\end{eqnarray}}

\newcommand{\bk}{{\bf k }}

\newcommand{\bq}{{\bf q }}

\newcommand{\beq}{\begin{equation}}
\newcommand{\eeq}{\end{equation}}

\begin{document}
\title{Many-body effects and possible superconductivity in the 2D metallic surface states of 3D topological insulators}
\author{S. Das Sarma and Qiuzi Li}
\affiliation{Condensed Matter Theory Center, Department of Physics, University of Maryland, College Park, Maryland 20742}
\date{\today}
\begin{abstract}
We theoretically consider temperature and density-dependent electron-phonon interaction induced many-body effects in the two-dimensional (2D) metallic carriers confined on the surface of the 3D topological insulator (e.g. Bi$_2$Se$_3$). We calculate the temperature and the carrier density dependence of the real and imaginary parts of the electronic self-energy, the interacting spectral function, and the phonon-induced velocity renormalization, enabling us to obtain a simple density and temperature dependent {\it effective} dimensionless electron-phonon coupling constant parameter, which increases (decreases) strongly with increasing density (temperature). Our theoretical results can be directly and quantitatively compared with experimental ARPES or STS studies of the 2D spectral function of topological insulator surface carriers. In particular, we predict the possible existence of surface superconductivity in Bi$_2$Se$_3$ induced by strong electron-phonon interaction.
\end{abstract}

\pacs{72.80.Vp, 81.05.ue, 72.10.-d, 73.22.Pr}

\maketitle

Electronic properties of 2D metallic states on 3D topological insulator (e.g. Bi$_2$Se$_3$) surfaces\cite{kim2011cho,dohun_PRL12TIph,qzliculcer_PRB12TI} continue to be one of the most active areas of condensed matter physics and materials science research world-wide\cite{hasan2010,FKM_PRL07,xiaoliang_RMP11,Culcer_PhysicaE2011,BAberneivig}. The reason for this intense interest is partly fundamental and partly technological. The 2D metallicity of the surface carriers here is topologically protected by the time reversal symmetry\cite{kane2005} as long as the bulk is a gapped insulator\cite{xia2009}, and as such, it is a new kind of a 2D electron system where carrier back-scattering is completely suppressed thus making it possible for a 2D system to be a metal even at $T=0$ in the presence of arbitrary amount of impurity disorder\cite{Hsieh_Science09}. This is in sharp contrast to ordinary 2D electron systems which are generically insulators at $T=0$ in the presence of any finite disorder due to the complete destructive quantum interference arising from the $2 k_F-$scattering. The 2D metallic surface carriers in topological insulators form helical bands with massless linear Dirac like energy dispersion protected by time reversal invariance\cite{FKM_PRL07,moore2007}.

Since the basic phenomenon of 3D topological insulators (TI) with the associated protected 2D gapless metallic surface carriers arises entirely from purely single-particle physics, much of the theoretical work on the subject has focused on the single-particle aspects involving topology and symmetry, including band structure\cite{zhang2009ti}, magnetic impurities\cite{hehongtao_PRL11}, magneto-electric response\cite{essin2009,xlqi_Science09,Maciejko_PRL10}, integer quantum Hall effect\cite{leedunghai_PRL09}, localization\cite{haizhou_PRL11}, and topological classification\cite{schnyder2008,kitaev_AIP09}. In the current work we focus on many-body TI properties with a theoretical investigation of the interaction-induced renormalization of the single-particle properties of the surface 2D carriers. Such many-body renormalization of the single-particle properties manifest themselves directly in various spectroscopic measurements such as angle resolved photoemission spectroscopy (ARPES)\cite{park_PRB10,hatch_PRB11,panzh_PRL12} and scanning tunneling spectroscopy (STS)\cite{roushan2009}, and therefore, our theory presented in this work provides direct predictions for ARPES and STS studies of TI surface metallic states. Since our theory provides the renormalized quasiparticle energy dispersion and level broadening, it is of crucial importance in understanding many properties of the TI surface states.

In discussing electronic many-body effects in general, it is useful to distinguish between electron-electron and electron-phonon interaction effects. Both are, of course, always present in any experimental solid state system, and can, in fact, be studied on an equal footing theoretically since the electron-phonon interaction can be eliminated in favor of an effective phonon-mediated electron-electron interaction through a canonical transformation which simply adds to the direct electron-electron Coulomb interaction to produce a complicated effective electron-electron many-body interaction including phonon effects implicitly\cite{jalabertdas_PRB89phonon}. Such an effective Hamiltonian is a good starting point when electron-electron and electron-phonon interactions are of comparable strengths. 2D carriers on TI surfaces are, however, very weakly interacting Coulomb systems by virtue of the bulk material (i.e. Bi$_2$Se$_3$\cite{xia2009,Hsieh_Nature09}, Bi$_2$Te$_3$\cite{Chen07102009,hsieh2009}, Bi$_{1-x}$Sb$_x$\cite{hsieh2008}, etc.) usually having a very large lattice dielectric constant ($> 50$) which strongly suppresses the effective 2D Coulomb-interaction (going as $2 \pi e^2 / \kappa q$ where $q$ is the 2D momentum transfer with $\kappa$ the background lattice dielectric constant). The Coulomb interaction strength (i.e. the ratio of the Coulomb potential energy to the noninteracting kinetic energy) in the 2D surface carriers, defined by the effective fine-structure-constant $r_s = e^2/(\kappa \hbar v_F)$ where $v_F$ is the velocity (typically $v_F \approx 7 \times 10^7$ cm/s) of the linear, massless Dirac band dispersion of the 2D carriers, is very small ($r_s \sim 0.05$) by virtue of the large background lattice dielectric constant. As such all effects of direct electron-electron Coulomb interaction can be safely neglected in discussing TI electronic properties except perhaps at extremely small 2D carrier density of little experimental or technological interest\cite{davidabergel_PRB13}.

\begin{figure}[htb]
\begin{center}
\includegraphics[width=0.99\columnwidth]{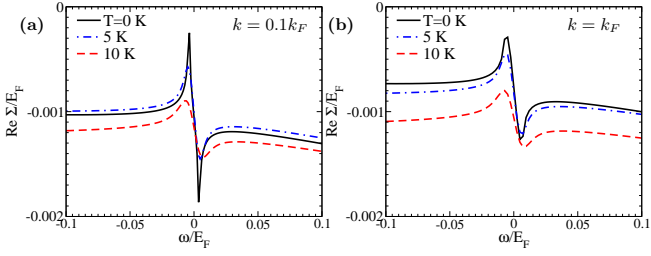}
\caption{ $\text{Re} \Sigma(k,\omega)$  of the surface state of Bi$_2$Se$_3$ as a function of $\omega/E_F$  for $n=3 \times 10^{12}$ cm$^{-2}$. (a) $ k=0.1 k_F$. (b) $ k= k_F$.
}
\label{fig:1}
\end{center}
\end{figure}

The starting point of our theory is the calculation of the finite-temperature electron self-energy arising from the electron-phonon interaction, which can be written in the leading-order theory as\cite{Grimvall}
\begin{align}
&\text{Re} \Sigma (k,\omega) = \sum_{s' = \pm 1}\int \frac{d^2 q}{(2 \pi)^2} |M|^2 \frac{1+  s' \cos\theta}{2} \nonumber
\\
&\times \Big[ \frac{n_F(\epsilon') + N_0 (\omega_q)}{\omega + \omega_q - \epsilon'} + \frac{1- n_F(\epsilon') + N_0 (\omega_q)}{\omega - \omega_q - \epsilon'}\Big]
\label{eq1:re}
\end{align}
and,
\begin{align}
\text{Im} \Sigma (k,\omega)& = - \pi \sum_{s'=\pm 1} \sum_{\nu=\pm 1}\int \frac{d^2 q}{(2 \pi)^2} |M|^2 \frac{1+  s' \cos\theta}{2} \nonumber
\\
&\times \left[n_F(\omega_q + \nu \omega) + N_0(\omega_q)\right]\delta(\omega + \nu \omega_q - \epsilon')
\label{eq2:im}
\end{align}
In Eqs.~\eqref{eq1:re}, \eqref{eq2:im}, $\Sigma = \text{Re} \Sigma+ i \text{Im} \Sigma$ is the wavevector or momentum ($k$) and energy or frequency ($\omega$) dependent finite-temperature 2D electron self-energy arising from the interaction of the 2D carriers with the effective TI surface acoustic phonons with the phonon dispersion given by $\omega_q = \hbar v_l q$ where $v_l$ is the surface phonon velocity. The electron-phonon interaction is considered to be mediated by the deformation potential coupling\cite{thalmeier_PRB11} quantified by the coupling strength $M$ given by: $|M|^2 = \frac{D^2 \hbar q}{ 2 \rho_m v_l}$ with $D$ and $\rho_m$ respectively being the deformation potential coupling strength and the 2D (ionic) mass density of the TI material. In Eqs.~\eqref{eq1:re} and \eqref{eq2:im}, $\epsilon' \equiv s' \hbar v_F |\bk + \bq| - \mu$ denotes the linear massless 2D energy dispersion of the surface metallic electron-hole bands (with $s' = \pm 1$ denoting electrons/holes in the conduction/valence band) measured with respect to the finite-temperature chemical potential $\mu$, and $n_F(\epsilon')$ is the finite-temperature Fermi distribution function corresponding to the carrier energy $\epsilon'$. Finally, $N_0(\omega_q)$ is the finite-temperature Bose distribution function for phonons with energy $\omega_q$. We mention that the $\frac{1+ s' \cos\theta}{2}$ factor is a matrix element effect arising from the topological nature of the surface bands which suppresses backscattering (i.e. $\cos \theta = -1$ for $\theta = \pi$) due to the chiral nature of the surface 2D bands associated with the intrinsic spin-orbit coupling in the system. We note that although we explicitly consider 2D surface phonons, the same theory should also apply for the interaction of the surface 2D electrons with 3D bulk acoustic phonons where a sum over all the bulk transverse phonon wavevectors have to be carried out, thus reducing the bulk phonon coupling to an effectively 2D coupling problem any way. We do not, however, believe that the coupling to bulk phonons would play a significant role in the physics under consideration since our interest here is restricted entirely to the surface 2D carriers.

\begin{figure}[htb]
\begin{center}
\includegraphics[width=0.9\columnwidth]{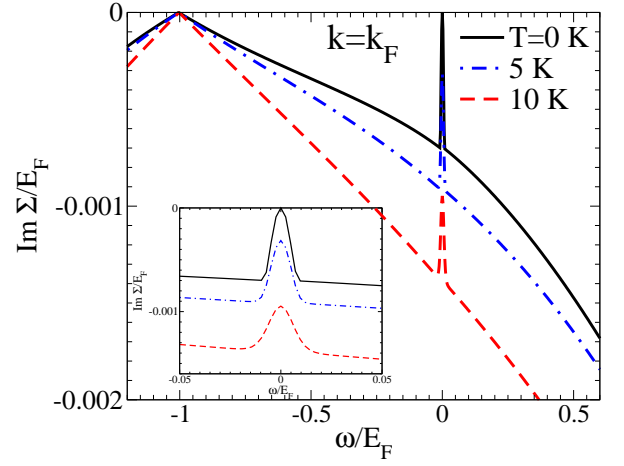}
\caption{ $\text{Im} \Sigma(k,\omega)$  of the surface state of Bi$_2$Se$_3$ as a function of $\omega/E_F$  for $n=3 \times 10^{12}$ cm$^{-2}$ and $k= k_F$. The inset presents $\text{Im} \Sigma(k,\omega)$ around $\omega =0$.
}
\label{fig:2}
\end{center}
\end{figure}

The finite-temperature electron-phonon self-energy defined by Eqs.~\eqref{eq1:re} and \eqref{eq2:im} arises directly from the formal expression $\Sigma \sim \int G |M|^2 P$ for the self-energy where an integration over all internal energy/momentum is implied with $G, P, M$ being the electron propagator, the phonon propagator, and the electron-phonon interaction matrix element respectively. This leading-order expression for the self-energy is, in fact, essentially exact for the weak-coupling problem by virtue of Migdal's theorem guaranteeing all vertex correction contributions being parametrically small in powers of $v_l/v_F$ ($\lesssim 0.01$ in the current problem).

Once the electron self-energy is calculated using Eqs.~\eqref{eq1:re} and \eqref{eq2:im}, there are two possible alternative (and, in principle, inequivalent) methods to define an effective dimensionless electron-phonon coupling strength parameter '$\lambda$' for the system (which depends on the carrier density $n$ and temperature $T$) given by\cite{Grimvall,PBAllen_PRB70}
\begin{equation}
\lambda_r \equiv  - \frac{\partial\text{Re} \Sigma (k_F,\omega)}{\partial \omega}\Bigg|_{\omega = \xi_{k_F}}
\label{eq4:lambd}
\end{equation}
where $\xi_{k_F} =  \hbar v_F k_F - \mu$, and\cite{PBAllen_PRB71}
\begin{equation}
\lambda_i \equiv \frac{- \text{Im} \Sigma (T_F \gg T \gg T_{BG})}{\pi k_B T} \Bigg|_{\omega = \xi_{k_F}}
\label{eq5:lamim}
\end{equation}
where $T_F=E_F / k_B = \hbar v_F k_F /k_B$, $T_{BG} = 2 \hbar k_F v_l / k_B$ are respectively the Fermi temperature and the Bloch-Gr\"{u}neisen temperature with $k_F \equiv (4 \pi n)^{1/2}$. The effective coupling constant $\lambda_r$ in Eq.~\eqref{eq4:lambd} directly provides the (density and temperature dependent) 2D carrier velocity renormalization due to the electron-phonon interaction\cite{PBAllen_PRB71}: $v_F^* = v_F (1+\lambda_r) ^{-1}$. The effective coupling constant $\lambda_i$ in Eq.~\eqref{eq5:lamim} is the high-temperature electron-phonon coupling parameter directly contributing to the phonon-induced carrier resistivity for $T \gg T_{BG} = 2 \hbar v_l k_F / k_B$ --- in particular, $\rho_T \propto \lambda_i T$ for $T \gg T_{BG}$, where $\rho_T$ is the phonon-dominated  linear-in-temperature electronic resistivity at high temperatures above the Bloch-Gr\"{u}neisen regime\cite{Giraud_PRB11,dohun_PRL12TIph}.

Before presenting our detailed results for the electron-phonon self-energy, we mention that the standard textbook definition of the dimensionless electron-phonon coupling constant in metals (e.g. the coupling strength entering the standard BCS-Eliashberg theory for phonon-induced superconductivity) is based on the electronic density of states (DOS) at the Fermi energy and is defined as\cite{PBAllenEinenkel_PR69}:
\begin{equation}
\lambda_d \equiv \frac{D_0(E_F) |M(q)|^2}{\omega_q}  \equiv \frac{ D^2 k_F}{4 \pi \hbar v_F \rho_m v_l^2}
\label{eq9:lam}
\end{equation}
where $D_0 (E_F) = k_F/(2 \pi \hbar v_F)$ is the electronic density of states of the surface 2D Dirac carriers at the Fermi surface ($k= k_F$ or $E = E_F = \hbar v_F k_F$).

How do $\lambda_r$, $\lambda_i$, $\lambda_d$ relate to (or compare with) each other? This is, in fact, a main topic discussed in this work, but it maybe useful to mention here that we find the three definitions to be completely consistent with each other in their respective regimes of validity.

Using the delta function in Eq.~\eqref{eq2:im}, and using the high-temperature ($T \gg T_{BG}$) expression for the Bose distribution function $N_0$, it is straightforward to calculate the asymptotic form for $\text{Im} \Sigma$ in the high-temperature ($T_{BG} \ll T \ll T_F$) limit of the equipartition regime for the acoustic phonons, obtaining
\begin{equation}
\text{Im} \Sigma (k, \xi_k) \approx -\frac{1}{2} \frac{k_F}{\hbar v_F} \frac{D^2}{2 \rho_m v_l^2} k_B T
\label{eq10:imhight}
\end{equation}
with $\xi_k = \pm \hbar v_F k - \mu$ denoting electron/hole energy. Eq.~\eqref{eq10:imhight} combined with the definition for $\lambda_i$ given in Eq.~\eqref{eq5:lamim} leads to $\lambda_i = \frac{k_F}{2 \pi \hbar v_F} \frac{D^2}{2 \rho_m v_l^2}$, which is exactly the same as the DOS definition for the coupling strength $\lambda_d$ given in Eq.~\eqref{eq9:lam}. Since $\lambda_i$ also gives the phonon scattering contribution to the high-temperature carrier resistivity, we conclude that the DOS and the transport definition of the electron-phonon coupling parameter $\lambda$ agree with each other, i.e. $\lambda_d \equiv \lambda_i \equiv \lambda_t$ where $\lambda_t$ is the coefficient of the phonon contribution to the temperature dependent electronic resistivity for $T \gg T_{BG}$ where the electronic resistivity is linear in $T$\cite{dohun_PRL12TIph}.

\begin{figure}[htb]
\begin{center}
\includegraphics[width=0.99\columnwidth]{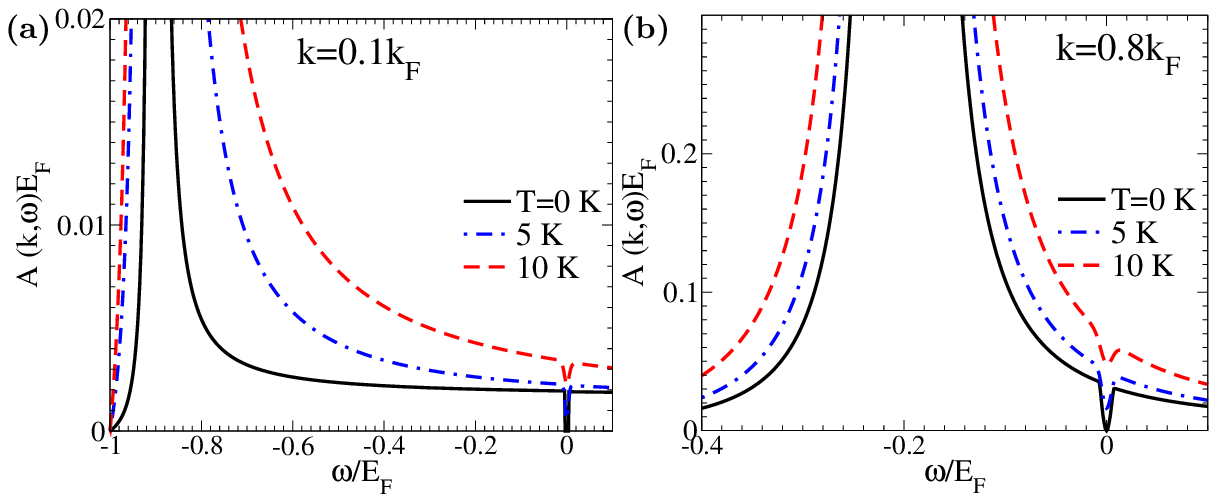}
\caption{ $A(k,\omega)$  of the surface state of Bi$_2$Se$_3$ as a function of $\omega/E_F$  for $n=3 \times 10^{12}$ cm$^{-2}$. (a) $ k=0.1 k_F$. (b) $ k= 0.8 k_F$.
}
\label{fig:3}
\end{center}
\end{figure}

The velocity renormalization factor $\lambda_r$ defining the many-body suppression of the carrier Fermi velocity\cite{PBAllen_PRB71}, $v_F^*/v_F = (1+\lambda_r)^{-1}$, is obtained from the real part of the electron-phonon self-energy defined in Eqs.~\eqref{eq1:re} and \eqref{eq4:lambd}. At $T=0$, the frequency derivative at the Fermi energy in Eq.~\eqref{eq1:re} can be analytically evaluated to obtain $\lambda_r = \frac{k_F}{2 \pi \hbar v_F} \frac{D^2}{2 \rho_m v_l^2}$, which agrees precisely with the DOS definition of the coupling parameter. Having established the complete equivalence among the definitions of the three distinct electron-phonon coupling strength parameters ($\lambda_d$, $\lambda_i$, and $\lambda_r$), we now focus on the electron-phonon self-energy function at finite temperature and carrier density by directly numerically integrating Eqs.~\eqref{eq1:re} and \eqref{eq2:im}. The corresponding electronic spectral function $A (k, \omega)$ is given by $A(k, \omega) \equiv - 2 \text{Im} G(k, \omega)$,  where $G(k,\omega) = [\omega - \xi_k - \Sigma(k,\omega)]^{-1}$ is the renormalized propagator for the 2D surface electrons including electron-phonon interaction.

In Figs.~\ref{fig:1}, \ref{fig:2} and \ref{fig:3}, we show our representative numerical results for $\text{Re} \Sigma$, $\text{Im} \Sigma$ and $A$, respectively. For numerical calculations, we use  the deformation potential $D=22$ eV, $v_F = 7 \times 10^{5}$ m/s, the mass density of Bi$_2$Se$_3$ for a single quintuple layer $\rho_m \simeq 7.68 \times 10^{-7}$ g/cm$^{-2}$, and the surface phonon velocity $v_l = 2900$ m/s\cite{dohun_PRL12TIph}. In addition to the well-defined quasiparticle peak, there is substantial background incoherent contribution to the spectral weight and a sharp (but very small) zero-energy feature associated with the carrier coupling to the acoustic phonon mode.

\begin{figure}[htb]
\begin{center}
\includegraphics[width=0.99\columnwidth]{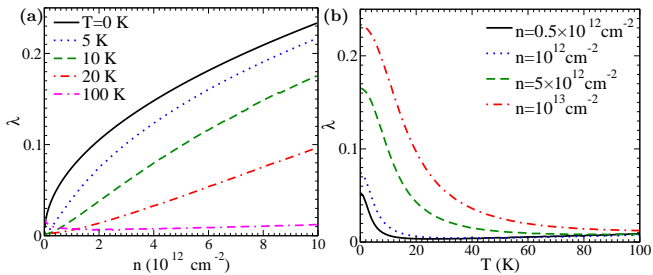}
\caption{ (a) $\lambda_r$  as a function of carrier density for different temperatures $T=0, \ 5, \ 10, \ 20, \ 100$ K from top to bottom. (b)  $\lambda_r$  as a function of temperature for different carrier densities $n=0.5$, $1$, $5$ and $10 \times 10^{12}$ cm$^{-2}$ from bottom to top.
}
\label{fig:4}
\end{center}
\end{figure}

Our most important results are shown in Fig.~\ref{fig:4}, where we show our calculated electron-phonon coupling parameter $\lambda$ ($\equiv \lambda_r$), calculated on the basis of Eq.~\eqref{eq1:re} through a direct numerical differentiation of the self-energy function. Since the quasiparticle velocity is given by $v_F^* = v_F (1+\lambda)^{-1}$, Figs.~\ref{fig:4}(a) and \ref{fig:4}(b) directly provide, as a function of density and temperature, the experimentally relevant carrier velocity renormalization by the electron-phonon interaction. It is interesting to note the very strong (weak) density (temperature) dependence of the coupling parameter for low (high) temperature (density), respectively. Thus, while $\lambda$ at $T=0$ varies in Fig.~\ref{fig:4}(a) as $\sqrt{n}$ with carrier density (consistent with the $\lambda \propto k_F$ analytical behavior), the $\lambda$-parameter at $T=100$ K in Fig.~\ref{fig:1} varies very little with carrier density (being very small, $\lambda \sim 0.01$ throughout). Similarly, we see in Fig.~\ref{fig:4}(b) that $\lambda(T)$ for $n = 10^{13}$ cm$^{-2}$ (relatively high density) starts at a high value ($\lambda \sim 0.25$), but quickly falls to around $0.05$ for $T \sim 40$ K.

Although all our calculations are done using the parameters for the Bi$_2$Se$_3$ TI system, which is by far the most extensively studied TI system in the literature, we believe that the general trends we find in our work (e.g. the functional dependence of $\lambda (n, T)$ on density and temperature) should be valid for 2D surface states on all TI materials. In this context, it is crucial to emphasize the obvious fact that $\lambda \propto D^2$, and as such all our quantitative results depend on our choice of $D=22$ eV for Bi$_2$Se$_3$ system, which we take from the recent detailed transport measurements\cite{dohun_PRL12TIph} of the temperature-dependent resistivity of the Bi$_2$Se$_3$ surface 2D carriers, which then lead directly to an estimate of $\lambda_i$ and hence of $D^2$. We emphasize that the subject matter of the electron-phonon coupling strength in the context of the Bi$_2$Se$_3$ surface metallic carriers\cite{hatch_PRB11} has been highly controversial with claims ranging from very weak coupling ($\lambda < 0.1$)\cite{panzh_PRL12} to very strong coupling ($\lambda  \gtrsim 0.5$)\cite{zhuxuetao_PRLTIphonon}. Our work should  resolve this controversy because, as our Figs.~\ref{fig:4}(a) and \ref{fig:4}(b) clearly show, the electron-phonon coupling parameter $\lambda$ in the Bi$_2$Se$_3$ surface 2D states depends very strongly on the temperature and carrier density. Since $\lambda$ increases strongly with carrier density (Fig.~\ref{fig:4}(a)) particularly at lower temperatures ($ < 20$ K), the coupling can be increased arbitrarily by increasing carrier density and could reach as high as $\lambda \sim 0.3 - 0.4$ for $n \gtrsim 10^{13}$ cm$^{-2}$. By contrast, $\lambda$ decreases sharply with temperature (Fig.~\ref{fig:4}(b)), and thus even at high carrier density ($n \sim 10^{13}$ cm$^{-2}$), the coupling constant will be very small ($\lesssim 0.01$) for $T \gtrsim 100$ K. A direct observation of the strongly temperature and density dependent quasiparticle velocity renormalization, as predicted in our Figs.~\ref{fig:4}(a) and \ref{fig:4}(b), could completely settle the controversy about the precise value of $\lambda$ (and hence $D$) in Bi$_2$Se$_3$.  We emphasize that the electron-phonon coupling in the 2D surface states of Bi$_2$Se$_3$ is much stronger ($\lambda \sim 0.25$) quantitatively than the corresponding coupling in graphene ($\lambda \lesssim 0.1$)\cite{tsepark_PR07} and 2D GaAs ($\lambda \lesssim 0.01$)\cite{kawamuradasminhwang_PRB} systems for comparable temperatures and carrier densities.

Since the original definition of the electron-phonon coupling parameter $\lambda$ (as used in our work) arose in the context of the BCS superconductivity (and the associated electron-phonon interaction in 3D metals), where $\lambda$ is indeed a constant for each metal, it may appear strange that we are discussing here (Figs.~\ref{fig:4}(a) and \ref{fig:4}(b)) a $\lambda$-parameter which depends strongly with temperature and density. This seeming paradox is  resolved by realizing that in regular 3D metals, the carrier density is very high (corresponding to a Fermi energy of $\sim 10$ eV) which cannot be varied at all except for small changes in going from one metal to another. Thus, the 3D metals are essentially always in the high-density regime. Equally importantly, the relevant temperature scale for 3D metals is the Debye temperature ($T_D$) which is high $T \gtrsim 300$ K. Thus, any temperature dependence of the $\lambda$-parameter in 3D metals can only manifest itself at rather high temperatures $T \gg T_D$, which perhaps does happen as reflected in the so-called ``resistivity saturation" phenomenon. In our 2D metallic system on the TI surfaces, however, the electron temperature scale ($T_F \propto \sqrt{n}$) and the phonon temperature scale ($T_{BG} \propto \sqrt{n}$) are both relatively small, leading to a very strong temperature dependence of the electron-phonon coupling ``constant". (As an aside we mention that the relevant phonon temperature scale is either $T_{BG}$ or $T_D$ depending on whichever is smaller--in 3D metals $T_{BG} \sim 10^4$ K $\gg T_D$ since $k_B T_{BG} = 2 \hbar v_F k_F$ is a very large energy scale because of the very high value of $k_F$ in metals, on the other hand $T_{BG} \ll T_D $ in 2D systems since $k_F \propto \sqrt{n}$ is typically small.)

We conclude by mentioning an important consequence of our theoretical findings. Given the large value of $\lambda$ (at least for $n > 10^{12}$ cm$^{-2}$) and very weak Coulomb repulsion in the 2D TI surface states on Bi$_2$Se$_3$ we predict the possibility of phonon-induced surface BCS superconductivity in Bi$_2$Se$_3$. In fact, using our calculated $\lambda$ (and taking the Coulomb repulsion parameter $\mu^* =0$) we predict $T_c \gtrsim 1$ K for Bi$_2$Se$_3$ surface 2D states for $n \gtrsim 5 \times 10^{12}$ cm$^{-2}$. The relevant mean-field $T_c$ formula here is easily obtainable from a BCS-Eliashberg theory, where $\lambda$ in our work is related to the Eliashberg function $\alpha^2 F(\omega)$ by $\lambda = \langle \alpha^2 F(\omega) \rangle_{\omega}$\cite{Grimvall}, which gives the following approximate $T_c$ expression for the surface 2D system: $T_c \approx T_D \text{exp}[-(1+\lambda)/(\lambda-\mu^*)]$. Using $T_D \approx 200$ K, $\lambda = 0.25$, and $\mu^* =0$, we get $T_c = 1.35$ K.  We propose that 2D superconductivity (with $T_c \sim 1$ K) should be looked for in the surface metallic states of Bi$_2$Se$_3$.

{\it Acknowledgements} This work is supported by ONR-MURI, LPS-CMTC and Microsoft Q.

\end{document}